# LUCID: A Framework for Reducing False Positives and Inconsistencies Among Container Scanning Tools

Md Sadun Haq[1], Ali S¸aman Tosun[2], and Turgay Korkmaz[1]

[1]*University Of Texas At San Antonio*
[2]*University Of North Carolina at Pembroke*

*Abstract*—

Containerization has emerged as a revolutionary technology in the software development and deployment industry. Containers offer a portable and lightweight solution that allows for packaging applications and their dependencies systematically and efficiently. In addition, containers offer faster deployment and near-native performance with isolation and security drawbacks compared to Virtual Machines. To address the security issues, scanning tools that scan containers for preexisting vulnerabilities have been developed, but they suffer from false positives. Moreover, using different scanning tools to scan the same container provides different results, which leads to inconsistencies and confusion. Limited work has been done to address these issues. This paper provides a fully functional and extensible framework named LUCID that can reduce false positives and inconsistencies provided by multiple scanning tools. We use a database-centric approach and perform query-based analysis, to pinpoint the causes for inconsistencies. Our results show that our framework can reduce inconsistencies by 70%. The framework has been tested on both Intel64/AMD64 and ARM architecture. We also create a Dynamic Classification component that can successfully classify and predict the different severity levels with an accuracy of 84%. We believe this paper will raise awareness regarding security in container technologies and enable container scanning companies to improve their tool to provide better and more consistent results.

## I. INTRODUCTION

Nowadays, containers [2] are the building blocks of the software industry. Containers are popular due to their portability and ease of integration with the CI/CD workflows, making it easier for developers to create and ship their products faster. Containers are also faster and more lightweight than Virtual Machines (VMs) [4] because they have no dedicated operating system. Instead, they share the host's operating system, resulting in faster boot times. Docker [3] container is one of the most widely used container-based technology today, with DockerHub [5] being the most popular publicly accessible registry for Docker containers. DockerHub also supports images across multiple operating systems (Windows, Linux) and architectures (AMD, x86_64, ARM, ARM64, etc.), giving developers flexibility in choosing their development system.

However, as containers share the hosts' operating system, any successful attack on the container can compromise the entire host system. The weak isolation between the container and the host can boost performance with trade-offs in security. An analysis done in 2020 [6] states that around two million public DockerHub images have **Critical** vulnerabilities. A recent zero-day vulnerability **Apache-Log4j** [31] was announced against the Apache **log4j** library, which took the software industry by storm. The extensive use of the log4j library in JAVA applications allowed malicious users to exploit this vulnerability by opening a reverse shell and causing irreparable damage. This vulnerability can extend to containers as libraries can be encapsulated within containers. To prevent such cases, container scanning tools can scan a container for pre-existing vulnerabilities. Existing container scanning tools such as Clair [46], Trivy [16], Anchore [1], and Snyk [11] are used by popular cloud vendors such as Google [30], Amazon [14], IBM [33] and Microsoft [43] to secure their container ecosystems. But it is not clear how they combine and use multiple scanning tools to safeguard their container systems when research done by Berkovich et al. [13], and others show that using different tools for the same image produces different sets of results.

Inconsistencies among different security tools have been historically a significant problem. Not only has it been around for a long time, but it is also very difficult to find a solution to this problem that is satisfactory for everyone [19]. Different virus scanners, malware detection tools, and security scanning tools display results according to their algorithm, leading to inconsistent results and false positives. These problems stem from using a multitude of different databases to gather relevant information that is not consistent in the first place. Moreover, the algorithms used by each tool might be based on different data sources and even different technologies, making it difficult for an individual or organization to decide which tool should be used [25], [38], for securing its applications, workflow, and infrastructure.

Inconsistencies and false positives have become the unnecessary add-ons for any security scanning tool and need to be dealt with or at least reduced. Reducing these two metrics will not only provide a consistent view of the vulnerabilities within a container but also provide actual vulnerabilities present within a container.

In this paper, we propose a framework that can be used to detect and reduce the inconsistencies and false positives obtained when using multiple scanning tools to scan the same set of images.

Contributions of this paper are as follows:
- This paper proposes the first database-centric framework consisting of the output of multiple container scanning tools to perform fine-grained queries to investigate and find the root causes of false positives and inconsistencies.
- We reduce false positives by removing packages detected by the scanning tool but not present in the container. We reduce inconsistencies by comparing the base image type with the vulnerability assigner and its modification time and choosing the most recently updated severity from a credible assigner.
- We adjust labels for the vulnerabilities and develop a learning method that can dynamically classify and predict the correct severity rating of a vulnerability on the fly, with an accuracy of 84%.
- We propose an extensible framework that can overall reduce the inconsistencies and false positives by 70% brought forth because of using multiple scanning tools.

The rest of the paper is organized as follows. Section II has an overview of Related Work. In Section III, we describe the relevant Background. Then we talk about our Proposed Framework in Section IV. Section V talks about the importance of having a database within our framework and the queries performed to find the root causes of the inconsistencies. The Evaluation Environment and Experimental Results are explained in Section VI and Section VII respectively. Discussion and Limitations of the proposed work are provided in Section IX, and finally, we conclude with Section X.

## II. Related Work

In this section, we discuss related work on the investigation of inconsistencies in computer and container security, container packages, and machine learning detection for containers.

**Inconsistencies.** Dong et al. [27] conducted a comparative analysis of inconsistencies between CVE and NVD from January 1999 to March 2018. They found mismatches of around 40%. Jiang et al. [36] evaluated the inconsistencies of open source vulnerability in different portions of the infrastructure like Networks, Data centers, IoT systems, etc., and found that most of the inconsistencies are based on network-related components. Croft et al. [23] performed large-scale analysis of Software Vulnerabilities (SVs) and tried to pinpoint the causes of inconsistencies such as *age of the vulnerability, fix time for the vulnerability and vulnerability verification from multiple sources*. Woo et al. [54] developed V0Finder, which focused on finding the correct origins of the vulnerabilities reported in public and open-source communities. Anwar et al. [15] conducted an analysis of data in NVD and looked for the root causes of inconsistencies. Their results showed that migration of the scoring mechanism from V2 to V3, confusing naming conventions, and lag in updates were the main contributors to inconsistency. Cui et al. [24] implemented VulDetector with the help of a weighted feature graph. It is based on vulnerability-sensitive keywords and graph slicing to accurately pinpoint vulnerabilities in software programs. Islam et al. [34] developed a classification model, that looked for inconsistencies in source code.

**Container Security.** Shu et al. [50] investigated the vulnerabilities present in DockerHub for both official and community-based images. He found that the images, on average, have over 180 vulnerabilities, some of which have not been updated for 100 days. Nannan Zhao et al. [59] found that only 3% of the files in DockerHub are unique, meaning that if the vulnerabilities are not fixed in the source images, they can propagate from parent to child as docker images are built on top of another as layers. Shay Berkovich et al. [13] created a benchmark to evaluate scanning tools with the help of Anchore, Clair, Trivy, and a binary scanner. They scanned the top three most downloaded images from DockerHub, and their findings concluded that each scanning tool generated a slightly different vulnerability set for each image, proving that no scanning tool is perfect and that different scanning tools provide better results over a subset of images. Kaiser et al. [37] conducted security and performance evaluation of different container technologies for ARM architecture. Maruszczak et al. [41] analyzed distro-less-based images and argued that smaller images might have unsuspecting security flaws. Kwon et al. [39] implemented DIVDS framework that checked a container for vulnerabilities when uploading and downloading images to and from DockerHub. Zhu et al. [60] proposed Lic-Sec, an enhanced AppArmor Docker security profiler that combines Docker-Sec [40] and a modified version of LicShield [7]. Lic-Sec is tested against 40 real-world exploits, including denial of service and privilege escalation attacks. Javed et al. [35] used three scanning tools and argued that they provide inaccurate information and vulnerable OS packages are more widespread.

**Container Packages.** Zerouali et al. [55], [56] studied how a lack of regular updates and third-party packages such as Python, Node, and Ruby can affect vulnerabilities in DockerHub. Kelly et al. [20] used two different tools and a virus scanner to look for malware-like behavior among containers. Haq et al. [32] extensively analyzed official images in DockerHub with multiple scanning tools on both AMD and ARM platforms. They also concluded that most vulnerabilities come from Python and GCC-based packages. Katrine et al. [53] argued that certified images are the most vulnerable, along with Python and JavaScript languages.

Existing works use scanning tools to detect preexisting vulnerabilities. Because using a single tool might lead to false negatives, the authors suggest using multiple scanning tools [13], [20], [32], [32], [35], [55]. However, none of the mentioned works address the inconsistencies and false positives caused by multiple scanning tools. To the best of our knowledge, there is no prior work investigating the false positives and inconsistencies in the security ratings of container scanning tools. For this reason, we have developed an extensible and dynamic framework, LUCID, for detecting and removing the inconsistencies and false positives associated with using multiple scanners in cloud, edge, and IoT container systems.

## III. Background

**Containers** Containers [2] are a virtualization technology that enables the packaging of code, packages, software, and other dependencies in a complete system. This makes the code portable, meaning the same code can be executed seamlessly on multiple platforms or environments. Virtual Machines [4] are also able to do this, but they ship with their own operating system, which makes them *bulky* and difficult to port and transfer among different machines across the internet.

**Static Scanning Dilemma.** Different container scanning tools exist to detect pre-installed vulnerabilities in a container,

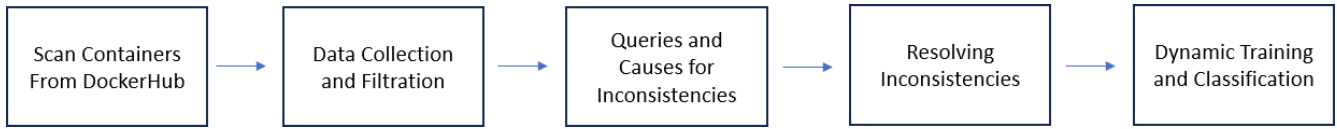

Fig. 1: Overview of LUCID Framework

but different vendors produce them, and they often produce different results for the same vulnerability, leading to inaccuracy and confusion. Moreover, in some cases, a scanning tool might pick up a vulnerability that does not exist in the first place, leading to false positives.

**Vulnerability Scoring Metrics.** Since containers are used to encapsulate different packages and software, bugs applicable to a specific software can apply to a container if the container in question contains the specific vulnerable software. In order to categorize existing vulnerabilities, the National Institute of Standards and Technology (NIST) [9] has introduced the CVSS scoring system, which provides different severity scores for different vulnerabilities, where each vulnerability is assigned a different CVE Identifier. The value of the score depends on multiple factors of the specific vulnerability and can change over time. The scoring system is of ascending nature, meaning the higher the score, the more severe the vulnerability. The scoring value ranges from 1.0-10.0 with None, Low, Medium, High and Critical severity ratings [44].

The CVSS Score Metrics are based on three metric groups: **Base, Temporal, and Environmental**.

The Base Metric contains the core characteristics of the vulnerability itself. These factors are independent of real-world exploits and patches provided by the industry to prohibit the specific exploit. Instead, they provide an initial starting point for understanding the vulnerability's true nature and the building blocks for patch development. The Base severity score, rating, and metrics can be found in the National Vulnerability Database (NVD) [10] tagged with their CVE Identifier. It is important to note that this score **does not change** over a period of time. The subcategories for the Base Metric Group are, Exploitability and Impact. Exploitability refers to how easy it is to exploit the vulnerability. Exploitability has further subcategories, including Attack Vector, Attack Complexity, User Interaction and Scope. These subcategories provide finer details, regarding the steps and effort required to perform an exploit. Impact refers to the effects of the vulnerability. Impact has further subcategories including Confidentiality Impact, Integrity Impact, Availability Impact, and Scope. These fields provide details regarding the full impact of the attack.

The Temporal Metric measures the current state of code availability and exploits techniques against the development of patches and fixes. This metric does **change over time**. The subcategories include Exploit Code Maturity, Report Confidence, and Remediation Level. Exploit Code Maturity refers to the current state of the exploit. Report Confidence measures a known vulnerability's degree of confidence and technical details. Remediation Level refers to the current patch state of the exploit.

The Environmental Metric allows organizations to modify the Base CVSS Score based on the industry's specific requirements. The subcategories are Security Requirements and Modified Base Metrics. Security requirements characterize the criticality of a vulnerability and its use case. Vulnerabilities on mission-critical systems tend to get a higher score when compared to ordinary systems. Modified base metrics enable researchers and analysts to adjust the Base Metrics according to the modifications within their personalized environment. For example, working on a server after making it unavailable to the internet may cause the Availability Impact to rise. However, this is misguided, as the server is purposefully taken down to install updates. Detailed information regarding all the categories can be found here [28].

Even though previous authors. [27], [20], [53], [32] and others talk about these inconsistencies and false positives in general, they do not provide a valid solution to this problem. In addition, obtaining the correct vulnerability score from multiple sources, especially from environmental, can be difficult and time-consuming. This is where our dynamic classification can help accurately predict and classify the vulnerability, with different metrics as its features. The framework's components are explained in detail in the later sections.

## IV. PROPOSED LUCID FRAMEWORK

Current scanning tools provide information regarding preexisting vulnerabilities in docker containers. However, the tools pull data from multiple sources that assign severities according to their algorithms, giving rise to inconsistency and confusion. Although it is difficult to reverse-engineer these algorithms, the causes for these inconsistencies can be traced back to the following candidates:

- **Inconsistent Package Name and Version**: The same vulnerability arises from different package names and versions with different severity ratings, giving multiple results for the same vulnerability.
- **Inconsistent Assigner**: The same vulnerability arises from different vulnerability assigners, giving different values. This is a major contributor, as the scanning tools used in the experiment had a total of 16 distinct assigners.
- **Inconsistent Modification Time**: The score for the same vulnerability changes due to it being modified over time. If the old scores are not removed, the same vulnerability appears with different severity ratings.

In order to address these problems, we present our framework, LUCID, given in Figure 1. We insert the data of the containers, scanning tools, and assigners in a database to look for entries with different severity ratings for the same vulnerability by performing queries. We then analyze which of the aforementioned candidates causes the inconsistency and try to resolve them. We also talk about the overhead and energy utilization of our framework. Now, we explain each member's functionality in detail.

**Scan Containers From DockerHub**. We first create a script that downloads all 168 official images from DockerHub during Feb 2023. We make use of the **V2** API provided by DockerHub. The API allows us to seamlessly surf DockerHub and download images using the keyword *Official* and looking for the most recent images. We then use the popular scanning tools, namely **Snyk**, **Clair**, and **Trivy**, to scan the downloaded Docker images. Scripts were created per tool so the container images could be scanned by the particular tool, and the resulting output would be appended to a JSON file.

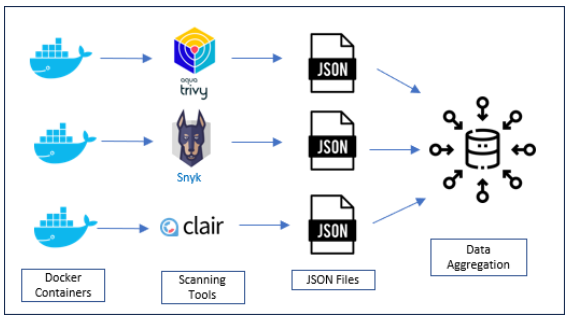

Fig. 2: Scanning of Docker containers and the aggregation of their results in JSON format.

**Data Collection & Filtration**. After obtaining the JSON files, appropriate data are filtered out, such as **cve_identifier, severity_rating, assigner**, etc., in order to correctly label the vulnerability. Clair has no modification_time nor assigner fields, so they were given NULL values. The combined data is then converted to a CSV file. Figure 2 shows the scanning of the containers and data aggregation process.

**Queries and Causes for Inconsistencies**. The CSV file is then used to enter data in a PostgreSQL database. Multiple queries are performed, and it is observed that the same CVE has multiple severity ratings. Additional queries, some of which are complex, are needed to narrow down the reasons for these inconsistencies. Some primary candidates for these inconsistencies are *mismatched package names and versions*, *different assigners and modification time*. For example, in order to find inconsistent assigners, we can look for vulnerabilities that have inconsistent severity and assigner for a matching cve_identifier pair. Figure 4 in Section V-A shows low-level details.

**Resolving Inconsistencies**: Queries related to the package details of the container are performed. Each scanning tool returns the vulnerability with a CVE identifier along with the name and version of the specific package. In addition, each container can be checked manually to see what packages are bundled within them. Package names and versions not present in the container but present in the results of the scanning tool are flagged as false positives. After detecting and reducing the false positives due to package mismatches, we try to reduce the inconsistencies that stem from different assigners and modification times. For this, we first download the data from the source of **Redhat, NVD, and Ubuntu**, at the same time we download images from DockerHub, as these are the primary assigners for most of the vulnerabilities. Different assigners might provide different severity ratings for the same vulnerability, so we prioritize the severity that matches the source and the OS of the container. Figure 6 in Section VII-A shows how we generate a consistent dataset by reducing false positives and inconsistencies.

**Dynamic Training and Classification**. After successfully removing and reducing the false positives and inconsistencies from our dataset, we implemented several supervised algorithms based on our revised data. We perform multi-class classification with Random Forest, Decision Tree, K-Nearest Neighbor, Gaussian Naive Bayes, and Multi-Layer Perceptron and analyze which algorithm performs best. Figure 9 in Section VIII shows how we use assigner data with the consistent dataset to train a model for dynamic classification.

V. QUERIES AND CAUSES FOR INCONSISTENCY

In this section, we explain why databases and queries are necessary to pinpoint the main causes of inconsistencies.

*A. Database and Queries*

A database allows us to perform an extensive analysis by looking for matching pairs and applying filters to pinpoint the exact causes of inconsistencies. The application of multiple filters allows us to see how the inconsistencies depend on certain criteria and on each other. Observing the results of the scanning tools in JSON format, we realized a need for a database and chose a popular relational database called PostgreSQL:14.0. We have experimented with a wide range of SQL queries to identify the underlying causes of inconsistencies and false positives.

**Tables and Structures.** A simple tabular structure was needed to store the results from our scanning experiment. The table and column names are simple and self-explanatory. The schema diagram shown in Fig 3 contains details about our tabular structure. The attributes that are used in **JOIN** operations are highlighted in the schema diagram. The table scan_results is used to store the initial results from scanning the container images with the three tools. The field package_type was picked up by the scanning tool, which best outlines the type of the image. It can have three values, Debian, Alpine, and Redhat. The field inner_modification_time is the last updated time of the vulnerability for each assigner. Snyk returns descriptive fields about the vulnerability, including a general last modification time and *modification times with specific assigners*. Clair had no Modification Time nor Assigner fields, so they were given NULL values.

We also create a table for each Assigner to track what severity the Assigner gives to the CVE and how and why they differ from each other and the scanning tools. However, as most of their fields match, we decided to display it as one table named assigner_results. The field cve_assigner holds the name of the Assigner, from which the data is pulled. The severity column is the severity rating given by the Assigner. This doesn't need to be consistent with other assigners. For example, Redhat gives its severity ratings where the values are Low, Moderate, Important, and Critical as opposed to NVD and Ubuntu which display Low, Medium, High, and Critical. Lastly, we also create a table that holds the information of an image and its package information. Because the scripts for adding data are reusable, any tool or assigner that needs to be added to the framework can be done quickly and easily.

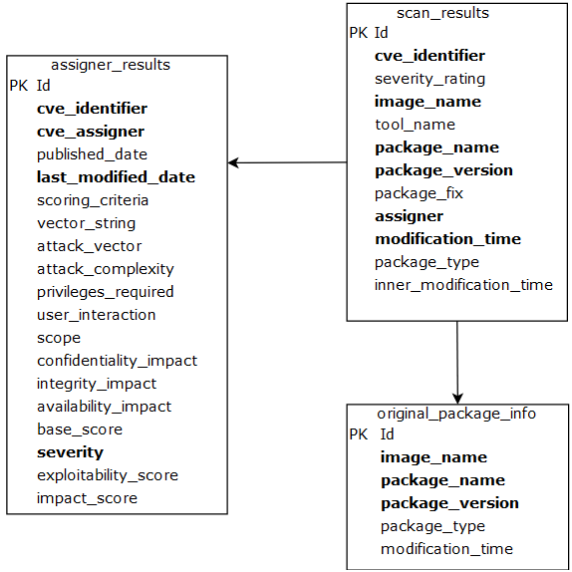

Fig. 3: Schema Diagram.

| Container Distribution | Package Commands |
|---|---|
| Ubuntu & Debian | dpkg -l |
| Redhat, CentOS & Fedora | rpm -qa |
| Alpine | apk info -vv |

TABLE I: Commands for different container distributions

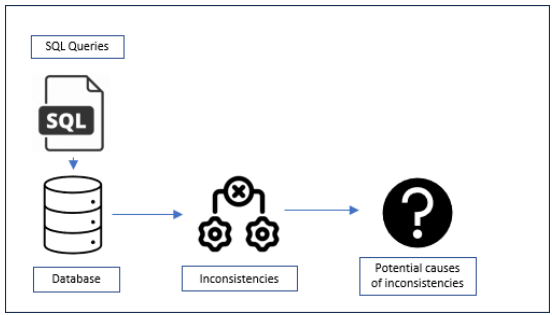

Fig. 4: Database and SQL queries and their role in detecting inconsistencies

**Queries.** With the database setup and data loading completed, we shifted our focus to writing queries. The queries we are interested in can be written easily using SQL. For example, to identify assigner inconsistencies, we find pairs of different assigners who report different severity ratings for the same vulnerability. The database allows this with the help of **JOIN** operations. For package version inconsistency, we find pairs of different package versions that report different severities for the same vulnerability and package name. Figure 4 shows the role of SQL queries to reduce the inconsistencies. To get the distribution of severities for all three tools, we use **GROUP BY** clause with tool name and severity rating. The GROUP BY clause is used to count and aggregate data and partition it with the given values that consist of tool name and severity rating. Fig 5 is created using the aforementioned GROUP BY clause. Some examples of inconsistencies are given below:

- **Package Name and Version Inconsistency**: We look for vulnerabilities with different severity ratings due to different package names. To look for inconsistent package versions, we do the same but assume that the package names match beforehand.
- **Assigner Inconsistency**: We look for vulnerabilities with different severities due to different assigners.
- **Modification Time Inconsistency**: We look for vulnerabilities with different severity ratings due to the same assigners but different modification time
- **Intra-Tool Inconsistency**: We look for vulnerabilities with different severity ratings due to the same assigners, modification time, and tool name.

We experimented with 138 queries initially, but the final framework worked with 42 queries. The initial queries were used to explore the causes of inconsistencies and pinpoint the root problems. The final set of queries was used to solve the inconsistencies.

### B. Root Causes Of Inconsistencies

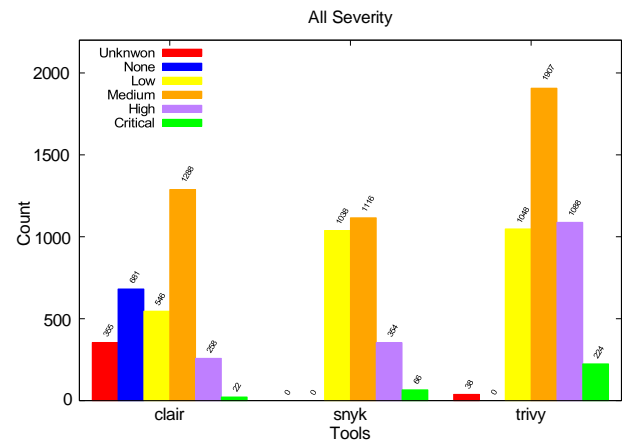

Fig. 5: Severity of Vulnerabilities

In this section, we discuss the leading causes of the inconsistencies among severity ratings for the same vulnerability. Figure 5 displays the severity distribution among the different tools for the same set of images. The tools do not agree with each other, which is a problem. We define the following to categorize inconsistencies.

**Package Inconsistency**: This inconsistency rises when the package name and/or version reports different severity ratings for the same vulnerability.

| Inconsistency Criteria | CVE | Severity Rating 1 | Candidate 1 | Severity Rating 2 | Candidate 2 |
|---|---|---|---|---|---|
| Package Name | CVE-2019-15847 | LOW | gcc-8 | HIGH | libmpx-2 |
|  | CVE-2020-8231 | LOW | curl | HIGH | libcurl4-openssl-dev |
|  | CVE-2022-0563 | LOW | util-linux | HIGH | libmount |
|  | CVE-2023-0288 | MEDIUM | vim | LOW | xxd |
| Package Version | CVE-2021-23840 | HIGH | curl:7.77.0.1-ph4 | MEDIUM | curl:7.68.0-1ubuntu2.6 |
|  | CVE-2022-48303 | MEDIUM | tar:1.30.5.3l8 | HIGH | tar:1.34-r0 |
|  | CVE-2023-0215 | HIGH | openssl:1.0.2k-19.amzn2.0.7 | MEDIUM | openssl:1.0.2k-25.el7_9 |
| Assigners | CVE-2021-4193 | MEDIUM | REDHAT | LOW | UBUNTU |
|  | CVE-2022-0213 | MEDIUM | REDHAT | LOW | NVD |
|  | CVE-2022-41903 | MEDIUM | UBUNTU | CRITICAL | NVD |
| Modification Time | CVE-2020-35527 | CRITICAL | 2022-12-08T22:29:00Z | MEDIUM | 2023-02-14T13:33:10.363339Z |
|  | CVE-2021-39537 | LOW | 2023-02-14T13:43:46.131616Z | MEDIUM | 2023-03-26T20:58:36.314633Z |
|  | CVE-2022-1292 | CRITICAL | 2023-02-14T15:08:45.939691Z | MEDIUM | 2023-03-27T02:04:36.826085Z |
| Intra-Tool | CVE-2021-33574 | CRITICAL | SNYK | LOW | SNYK |
|  | CVE-2021-35942 | CRITICAL | SNYK | LOW | SNYK |
|  | CVE-2021-3997 | MEDIUM | TRIVY | LOW | TRIVY |

TABLE II: Examples for Inconsistencies

For example, in Table II, we see package name and version mismatches. For CVE-2023-0288 we see Medium and Low severity ratings originating from vim and xxd packages respectively. For CVE-2022-48303 we see a package version mismatch that displays Medium and High severity ratings from different versions of tar package. Curl, OpenSSL, and Imagemagick mismatches of this category are present throughout all three tools. If the tool returns a vulnerability with a package that does not exist in the container, we call it **Hard False Positive Detection**. If the package names match, but the versions do not, we refer to it as **Soft False Positive Detection**. In order to find existing packages within a container, we employ commands from Table I, depending on the underlying container OS. Note this is not an actual OS, just an abstraction. Containers that do not ship with any OS are left out of this part of the experiment. We also try to track packages that are installed from the source. The best way to detect packages from the source is to locate them under the **/usr/local/bin** folder, where they are installed. Since installing directly from the source does not include any package manager, they do not appear in any of the commands shown in Table I. There were no entries via source, as we downloaded all the images directly from DockerHub. We conducted our experiment with package managers, as this is still a popular route for getting packages, and this returns packages that came with the container. We include the method for checking via the source in case users install packages from there. If installed from the source, it is necessary to check the package version from the /usr/local/bin/ folder or manual version lookup. Checking via the package manager might probably lead to incorrect results.

**Assigner Inconsistency**: This inconsistency arises when different **assigner** reports different severity ratings for the same vulnerability. Assigners are the public repository from which the tool pulled its data and cross-referenced the vulnerability. The assigners we consider are NVD, Redhat, and Ubuntu. NVD is the most popular source of information regarding any vulnerability, regardless of its origin. This is why NVD contains vulnerabilities from various platforms and provides a generalized view of the vulnerability landscape. On the other hand, Redhat and Ubuntu have their own security advisories that focus on vulnerabilities that emerge from their product. They provide a specialized view of the vulnerability that NVD might lack. This is also the reason why most vulnerabilities, in general, cannot be found in the Ubuntu and Redhat security advisories. In addition, in cases where NVD shows a severity that conflicts with the one provided by a specialized source, it is better to **prioritize** the severity rating provided by the **source** than NVD. Consider this example from [48], "*NVD may rate a flaw in a particular service as having a High Impact on the CVSS CIA Triad (Confidentiality, Integrity, Availability) where the service in question is typically run as the root user with full privileges on a system. However, in a Red Hat product, the service may be specifically configured to run as a dedicated non-privileged user running entirely in an SELinux sandbox, greatly reducing the immediate impact from compromise, resulting in Low impact*". The aforementioned statement signifies why a severity score provided by a specialized vendor can override that of a generic one. This is why we are also prioritizing the severity score and rating provided by the source vendor. Vulnerabilities that have multiple severity scores will be cross-checked with the **Container OS** of the image and that of the source. **The severity rating that will match with the source will be prioritized over all.** If matches among multiple severities with multiple sources exist, the match with the most counts shall override the others.

In Table II, CVE-2021-4193 has Medium and Low severity ratings for Redhat and Ubuntu assigners. CVE-2022-41903 also has Medium and Critical severities for Ubuntu and NVD assigners.

**Modification Time Inconsistency**: This type of inconsistency arises when different Modification Time, along with the same assigner, reports different severity ratings for the same vulnerability. Modification Time is the last time the record in the public repository was updated for the same assigner.

In Table II, CVE-2020-35527 has Critical and Medium severities, and CVE-2021-39537 has Medium and Low severity ratings for different timestamps, respectively.

**Intra Tool Inconsistency**: Intra-tool inconsistency or **Same Assigner and Same Modification Time** inconsistency exists when severity mismatches are present for the same CVE within the same tool. It can occur due to mismatches in package names, versions, assigners, and modification time. It can also occur even when all the criteria match, which we are showing here. It is unclear why this happens, as these inconsistencies

| Level | Inconsistency Type | Numbers | % |
|---|---|---|---|
| L1 | Inconsistent CVEs ALL | 1669 | 100.00% |
| L2 | Package Name Inconsistent CVEs | 791 | 49.10% |
| L3 | Package Version Inconsistent CVEs | 223 | 13.00% |
| L4 | Different Assigner Different Modification Time Inconsistent CVEs | 1591 | 95.00% |
| L5 | Same Assigner Different Modification Time Inconsistent CVEs | 960 | 57.50% |
| L6 | Same Assigner Same Modification Time Inconsistent CVEs | 301 | 18.00% |

TABLE III: Breakdown of Inconsistencies Top Down

| Level | Inconsistency Type | Numbers | % |
|---|---|---|---|
| L1 | Resolved Inconsistent CVEs ALL | 1183 | 70.1% |
| L2 | Resolved Package Name Inconsistent CVEs | 187 | 11.2% |
| L3 | Resolved Package Version Inconsistent CVEs | 28 | 1.68% |
| L4 | Resolved Different Assigner Different Modification Time Inconsistent CVEs | 6 | 0.36% |
| L5 | Resolved Same Assigner Different Modification Time Inconsistent CVEs | 661 | 39.60% |
| L6 | Resolved Same Assigner Same Modification Time Inconsistent CVEs | 301 | 18.00% |

TABLE IV: Resolving Inconsistencies Bottom Up

should not exist within the same tool.

In Table II, CVE-2021-33574 shows Critical and Low severities for Snyk, whereas CVE-2021-3997 displays Medium and Low Severity ratings for Trivy. Out of the three tools, Snyk suffers the highest from intra-tool inconsistency.

## VI. EVALUATION ENVIRONMENT

This section describes the evaluation environment, including the containers investigated, the scanning tools used, and the platform upon which experiments were conducted.

**Containers Investigated**. The experiment involved downloading 168 official images from DockerHub via the V2 API [26]. The API returns a multitude of images based on the Docker Image name. We chose the most updated version of the image from the list.

**Scanning Tools Used**. We used three popular open-source scanning tools to scan the containers. The tools returned results in a structured JSON format that included a combination of lists and dictionaries.

The tools return CVE (Common Vulnerabilities and Exposures) identifiers [8], a security flaw assigned a CVE number that can be referred to in the future. In addition, CVE identifiers are used with other database services like VulDB [12] and NVD (National Vulnerability Database) [9] that provide information about the vulnerability, severity ratings, impacts, and fixes. The severity ratings and scores are based on the CVSS metrics. The v3.0 is now the standard for measuring severity rating, whereas the previously employed v2.0 ratings are kept for historical purposes. Together, they provide a detailed view of the vulnerability and its probable impact. CVSS scoring metrics are given in Table **??**.

Below is a brief description of the tools used.
- **Trivy** [16] is an open-source comprehensive scanner, developed by Aqua. It has been created to scan images and detect vulnerabilities. Trivy is the easiest and simplest tool to use among the three.
- **Clair** [46] is another open-source project developed by Quay and RedHat. It looks for vulnerabilities by performing the static analysis of appc and Docker containers. Clair was the hardest of the three tools to set up and use. Its GitHub page had poor documentation, which resulted in many trials and errors.
- **Snyk** [11] is an enterprise-level security tool that provides scanning services for all platforms including containers. It had a free plan for one user, which we were able to use. Snyk was easier to install and use, compared to Clair, but needed a verification token to get started.

All the tools need to be installed on the host machine to perform scanning. The host machine had an Intel i7-2600 CPU with 3.40 GHz, along with 16GB RAM and 320GB of HDD, and the Ubuntu:20.04.6 LTS operating system.

## VII. EXPERIMENTAL RESULTS

In this section, we describe the experimental results in our evaluation environment.

### A. Results for Finding Inconsistencies

According to our queries performed, there are 3766 distinct CVEs found from scanning all the official images with the three scanning tools. We write **SQL** queries to find mismatches for the same vulnerability. Among these CVEs, 44% are mismatched (1669), making it 9.98 inconsistencies per image. Trivy had the highest number of inconsistencies, due to its higher number of data sources. The breakdown of these mismatches is given in Table III and the classification of the levels are given is Table V. More details are given below:

| Levels | Inconsistency Criteria |
|---|---|
| L1 | Everything is Inconsistent |
| L2 | Package Name is Inconsistent |
| L3 | Package Name is Consistent Package Version is Inconsistent |
| L4 | Assigner is Inconsistent |
| L5 | Assigner is Consistent Modification Time is Inconsistent |
| L6 | Assigner is Consistent Modification Time is Consistent Tool Name is Consistent |

TABLE V: Classification of inconsistencies according to different levels.

**Filtering False Positives**: We try to filter out false positives by trying to find package names and versions that are published by the tools, but are not present in the container. We were able to find such inconsistencies, but it did not reduce our list of inconsistent CVEs, as all of the False Positive candidates had **duplicate entries** that had matching package names. It seems that the tools sometimes label a vulnerability with **Unapproved** severity rating with the incorrect package name, but forget to remove the record when the same vulnerability is updated with the correct severity rating and package name, causing the same vulnerability to have two severity ratings. We removed these records before conducting our experiment but failed to reduce the inconsistent list of CVEs.

**Package Inconsistencies**: 49.10% (791) inconsistencies are present due to package name mismatches, whereas 13% (223) exist due to version mismatches. Take note, that for package version mismatches, we assume that package names are equal beforehand.

We find package name inconsistencies **L2** by finding pairs of package names and severity mismatches with the same image name. These image names are taken from our scanning results and getting package information by running the container. Since data gained from running the container does not contain any information on any vulnerability, we cannot use cve_identifier for matching. For package version mismatches **L3**, we look for mismatches in the package version and severity for the same pair of images and package names. Since we are adding an extra filter, package version inconsistencies are less than the package name inconsistencies.

**Assigner Inconsistencies**: Around 95% (1591) inconsistencies are present due to different assigners. We find these inconsistencies **L4** by finding matching pairs of cve_identifiers, but the assigners differ along with the severity rating. We exclude comparisons with package names and versions from here onward as the query becomes overly complex and time-consuming.

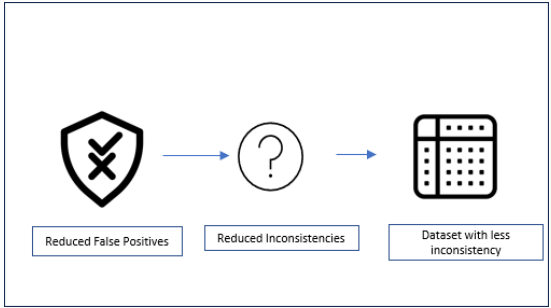

Fig. 6: Reducing False Positives and Inconsistencies to gain a less conflicted dataset .

| Algorithm | Abbr. | Parameters |
|---|---|---|
| Random Forest | RR | n_estimators=200, max_dpeth=32 |
| Decision Tree | DT | max_features=200 |
| K-Nearest Neighbor | KNN | algorithm='auto', leaf_size=30, metric='minkowski', metric_params=None, n_neighbors=5 p=2, weights='uniform' |
| Gaussian Naive Bayes | GNB | default |
| Multi-Layer Perceptron | MLP | solver='adam', alpha=1e-5, max_iter=1000,activation='relu', hidden_layer_sizes=(5, 2), random_state=1 |

TABLE VI: Parameters and abbreviations (Abbr.) for different algorithms

**Modification Time Inconsistencies**: Around 57.50% (960) inconsistencies are present due to the same assigner but different modification times.

We find these inconsistencies **L5** by finding matching pairs of cve_identifiers and assigners, but the modification time and severity rating differ.

**Intra-Tool Inconsistencies**: Around 18% (301) inconsistencies are present even if the assigner, modification time, and tool name are similar.

We find these inconsistencies **L6** by finding matching pairs of cve_identifiers, assigners, modification time, and tool names. The only thing different is the severity rating.

All the queries are elaborated with self-joins and unions to add results for Ubuntu, Redhat, and their severity_ratings compared. Self-joins are necessary to find inconsistencies within the same table. For example, if Snyk has multiple severity ratings for a vulnerability or if Redhat has two fields for one CVE. As each query is performed separately, there might be overlaps regarding certain CVEs.

*B. Results for Resolving Inconsistencies*

To resolve the inconsistencies, we use two approaches. To update the vulnerability with a more accurate severity rating, we propose the **Recent Approach**, as the most recently updated severity is expected to be the most accurate. The **Voting Approach** uses severity ratings matched with credible assigners and chooses the severity with the highest amount of votes. Further details are given below:

**Recent Approach**: If a single CVE has multiple entries, we take the severity with the most recent 'Modification Time' of the vulnerability, as that contains data that has been updated most recently. If multiple severity ratings exist for the same 'Modification Time', we move to the second approach.

**Voting Approach**: After we have chosen the most recent 'Modification Time' or when there is none, if multiple severity ratings exist for the same vulnerability, we choose the entry with the highest matching severity type.

Only L6 and L5 in Table IV will use both **Recent** and **Voting** approaches. L2, L3, and L4 will make use of the **Voting** approach only. For example, if CVE-2021-33574 has four entries with 'Low' and five entries with 'Medium' severity for the same 'Modification Time', we will choose Medium as the severity for CVE-2021-33574. For the rest of the levels, we will use the 'voting' approach.

To prioritize solving the inconsistencies, we use a **Bottom Up** approach. We will solve the inconsistencies from the bottom level and move to the top. As we correct the inconsistencies of L6, we record the CVEs that we have solved and exclude them from the inconsistencies of L5, we consider them already fixed. The logic behind this is that since L6 has inconsistencies that are detected by the same tool, have the same assigners, and most recent modification time, we rectify them only once. We don't need to rectify them again, as the severities on that level have already been fixed with the most consistent data. The same logic applies as we go up the levels from L5 to L1 in Table IV. L1 contains the total number of distinct CVEs that we were able to fix, 1183, which is around 70% of the mismatched CVEs.

In Table IV, we see the breakdown of fixes. We are able to fix all of the inconsistencies in L6. For L5 and L4, we are able to solve 39.6% and 0.36% of the mismatches separately. For L4, we apply the voting system, only if we are able to find matching package assigners that include 'Ubuntu', 'Redhat', or 'NVD'. For L3, we are able to fix 1.68% of the package version mismatches, and for L2, we are able to rectify 11.2% of the package name mismatches.

Overall, we were able to fix around 70% of the mismatches observed. Since each layer gave its separate fixed percentage, the values might be a bit skewed, as we are ignoring the CVEs that have been fixed in the layer below. However, the final value accounts for the total number of CVEs fixed across all five layers (from L2-L6).

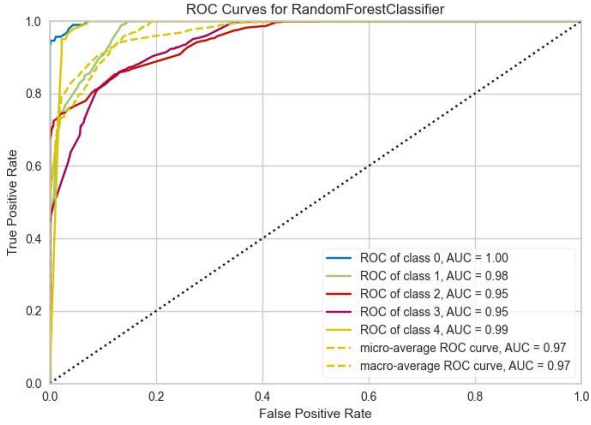

Fig. 7: MultiClass AUC ROC curve for Random Forest

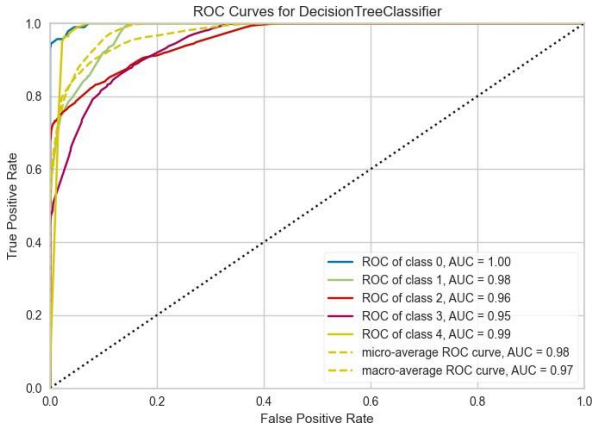

Fig. 8: MultiClass AUC ROC curve for Decision Tree

We cannot fix most inconsistencies from **L4**. This is because we do not apply the fix to those vulnerabilities that are reported from less reputable sources. We chose only three assigners out of thirteen different ones, as these three are the most credible, reputed, and shared information about the vulnerabilities in an open manner. Figure 6 shows the sub-component for reducing inconsistencies and false positives. We employ dynamic classification and use our improved dataset to train different models to better classify severities, especially those we were unable to fix.

## VIII. DYNAMIC CLASSIFICATION

After reducing the inconsistencies, we now have a dataset that has more consistent distribution of labels. Figure 9 shows cross-checking the dataset with NVD, Redhat, and Ubuntu security advisories, before applying machine learning to classify the different anomalies. With the data provided by the vulnerability assigner, we can choose the features that can be used to deduce the severity of a vulnerability. We develop a multi-classification problem with ten features and five possible outcomes. Our dataset was skewed, with 47.40% and 36.40% of the data belonging to the Medium and High classes, respectively. The remaining are split among None, Low and Critical, with each having 3.44%, 7.53%, and 5.21%, respectively. The input features are attack_vector, attack_complexity, privileges_required, user_interaction, scope, confidentiality_impact, integrity_impact, availability_impact, exploitability_score and impact_score. Numeric values were kept the same. For categorical variables, labelEncoder was used. The base_score is used to map a CVE to a severity using the CVSS V3 metrics present in Table **??**. The five severities are mapped as, **0-None**, **1-Low**, **2-Medium**, **3-High** and **4-Critical**. We first convert the base_score (continuous) to severity_rating (string) and then finally to numeric labels (discrete values). With this information, our model was able to predict the vulnerability's severity with the given features accurately. We perform a multi-class classification of the problem with different algorithms and analyze which algorithm performs best.

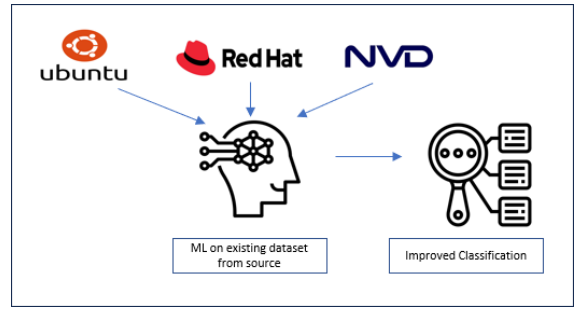

Fig. 9: Applying Machine Learning algorithm to dataset after cross-checking with source.

We implement Random Forest [21], Decision Tree [47], K-Nearest Neighbor [58], Gaussian Naive Bayes [57], and Multi-Layer Perceptron [29]. They are all imported from the scikit-learn library [45]. Their parameters can be found in Table VI. We use these algorithms because they are easy to implement and work well with small and imbalanced datasets. The parameters are chosen based on GridSearchCV [52]. We experiment with the most influential parameters with a wide range of values for each algorithm and choose the parameters that produce the most optimum results. We use 5-fold cross-validation for all the algorithms because that yields the best results. We display their ROC (Receiver Operating Characteristic) curves and AUC (Area under the ROC Curve) values as they display an aggregate measure of performance across True Positive and False Positive thresholds.

Their AUC values and ROC curves are plotted in Fig 7,8,10,11,12 separately, with the help of the library YellowBrick [18]. We use the **One-vs-Rest** approach where one class is plotted against all others (assumed as one). Due to the fact, that this is a multi-classification problem, it was not possible to include all five figures in one for the purpose of comparison. The comparison of AUC values for each class and algorithm is given in Table VII.

As we see from Fig. 7, Fig. 8 and Fig. 10, Random Forest, Decision Tree, and K-Nearest Neighbors perform relatively well, with the first two performing exceptionally. Table VII shows that the Decision Tree performs slightly better for class

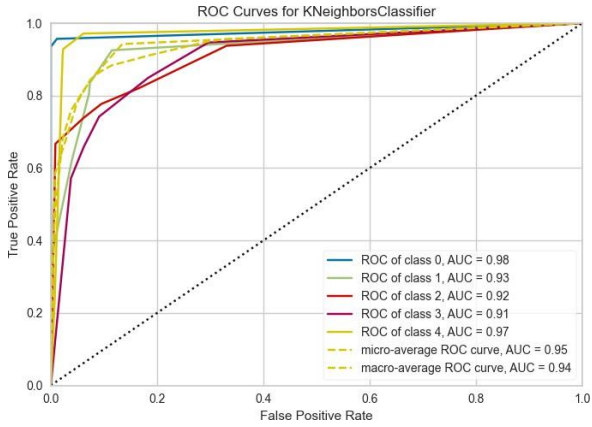

Fig. 10: MultiClass AUC ROC curve for K-Nearest Neighbors

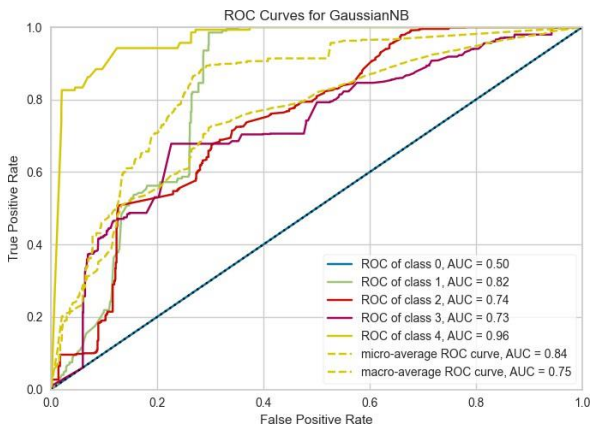

Fig. 11: MultiClass AUC ROC curve for Gaussian Naive Bayes

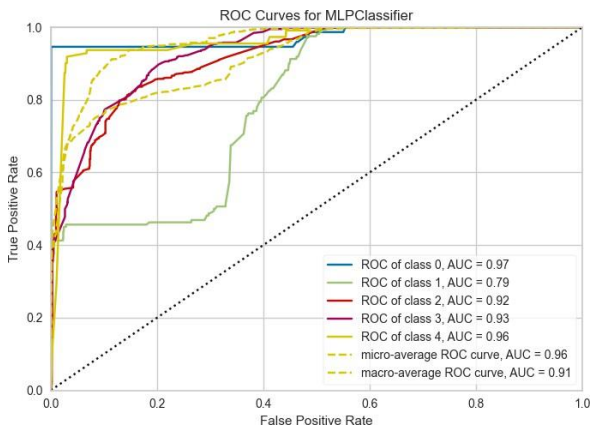

Fig. 12: MultiClass AUC ROC curve for Multi-Layer Perceptron

2. However, from Fig 11 and Fig 12, we see that Gaussian Naive Bayes and Multi-Layer Perceptron perform poorly. In fact, we see that from Fig 11 and Table VII, Gaussian Naive Bayes has an AUC score of 0.5 for class 0 and class 4, which is equivalent to random guessing. The AUC scores of classes 2 and 3 are also lower than others. Multi-Layer Perceptron performs better than Gaussian Naive Bayes but not as good as the other 3. According to Figure 12 and Table VII, it performs worst for class 1 with an AUC value of 0.79. Because ROC curves and AUC scores are plotted and calculated with the values of True Positive and False Positive only, we also calculate the Accuracy, Precision, Recall, and F1-Score for the algorithms so that we can obtain a complete view regarding the performance of each algorithm and provide an accurate comparison. We decided to use the Macro setting in calculating the metrics, as our goal is to evaluate performance across each class, not just focus on the detection rate. The Macro setting from scikit-learn [49] calculates the scores for each class separately and returns the *unweighted mean* across all five classes. It is vital to use unweighted because using weighted will shift the bias towards classes with more extensive data sizes and reduce the importance of classes with smaller data representation. The metrics can be found in Table VIII.

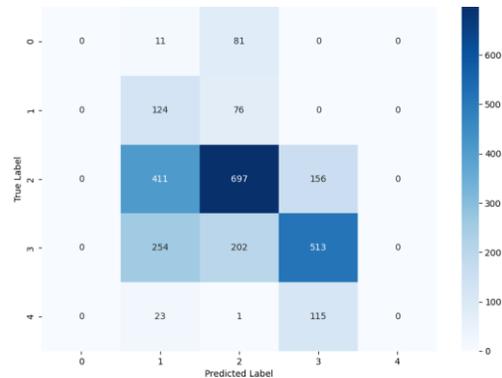

Fig. 13: Confusion Matrix for Gaussian Naive Bayes

We observe that the Decision Tree algorithm performs best with an Accuracy of 0.84 and an F1-score of 0.83, with Random Forest and K-Nearest Neighbors following closely. We also see that Multi-Layer Perceptron and Gaussian Naive Bayes have lower accuracy, with the latter having the lowest. Gaussian Naive Bayes also fall behind Multi-Layer Perceptron in all other metrics. Upon creating a confusion matrix in Fig 13 and closer inspection, we see that Gaussian Naive Bayes does not detect anything for classes 0 and 4. Classes 0, 1 and 2 have a very low amount of data. Gaussian Naive Bayes works well for classes 1,2 and 3, but due to no detection from the other two classes, it has abysmal Precision, Recall, and F1-Score. It was difficult to improve the performance of Gaussian Naive Bayes because its scikit-learn implementation [51] contains only two parameters to experiment with. In addition, repeating the experiment with multiple folds did little to improve its performance. Performance improvement due to the optimum number of folds likely stemmed from gaining a more uniform distribution of classes among the train and test partitions, which helped the model classify the data correctly. Increasing the number of folds further does not increase the performance of the model; instead, it only escalates the computation costs. The algorithms tend to assign higher severity scores for the incorrect classifications.

| Algorithm | 0 | 1 | 2 | 3 | 4 |
|---|---|---|---|---|---|
| Random Forest | 1.00 | 0.98 | 0.95 | 0.95 | 0.99 |
| Decision Tree | 1.00 | 0.98 | 0.96 | 0.95 | 0.99 |
| K-Nearest Neighbor | 0.98 | 0.93 | 0.92 | 0.91 | 0.97 |
| Gaussian Naive Bayes | 0.50 | 0.82 | 0.74 | 0.73 | 0.50 |
| Multi-Layer Perceptron | 0.97 | 0.79 | 0.92 | 0.93 | 0.96 |

TABLE VII: AUC values for each class and algorithm

Gaussian Naive Bayes obtains low scores for classes with low data, affecting its overall performance. Multi-Layer Perceptron performs a bit poorly for one of the classes, which causes its overall recall and f1-score to go down. This experiment highlights why it is necessary to use the ROC curve, AUC scores, and multiple metrics using *Macro* setting (due to skewed dataset) to measure the performance of a model.

### A. Resource Utilization

Due to the rise in Edge Computing and IoT devices, it is only a matter of time before these devices implement sophisticated learning models for the purpose of security, monitoring, and performance [17], [22], [42]. It is now the standard to release container images in DockerHub in ARM, ARM64 format along with its AMD and AMD64 counterparts. This is why we perform an analysis of the CPU and Memory Usages, made by the different algorithms. We do not perform any analysis for GPU as most of these devices, especially the inexpensive ones, do not come with it. In Fig 14 we see the mean CPU usage of all the algorithms. The abbreviations are mentioned in Table VI. We observe that Decision Tree and Gaussian Naive Bayes have the highest mean and standard deviation among the five algorithms. In Fig 15, we see that Random Forest, K-Nearest Neighbors, and Multi-Layer Perceptron have the highest average memory usage of 120 MB going up to a maximum of 180 MB. From this analysis, we can say that from a CPU and Memory perspective, Random Forest is the best algorithm suited to be used for Edge and IoT devices. Decision Trees and K-Nearest Neighbors are also good candidates, even though the former is more CPU-intensive. Taking both accuracy and performance efficiency into consideration, Random Forest is the best-suited algorithm for Dynamic Severity Classification in resource-constrained devices.

| Algorithm | Accuracy | Precision | Recall | F1 |
|---|---|---|---|---|
| Random Forest | 0.81 | 0.83 | 0.81 | 0.81 |
| Decision Tree | 0.84 | 0.83 | 0.83 | 0.83 |
| K-Nearest Neighbor | 0.79 | 0.78 | 0.81 | 0.78 |
| Gaussian Naive Bayes | 0.51 | 0.36 | 0.42 | 0.36 |
| Multi-Layer Perceptron | 0.77 | 0.85 | 0.57 | 0.59 |

TABLE VIII: Metrics for different algorithms in MACRO setting

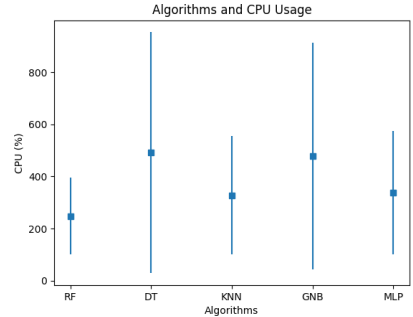

Fig. 14: Mean and Standard Deviation of CPU Usage

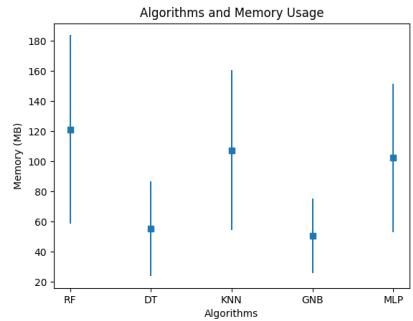

Fig. 15: Mean and Standard Deviation of Memory Usage

### B. LUCID Adoption and Overhead

We plan to deploy LUCID in the cloud, where users can upload their containers and have them scanned. As, LUCID has multiple components, it is essential to know the overhead of each component and that of the overall framework.

Since LUCID depends on different scanning tools, performance impairment or enhancement on their end will impact LUCID directly. Data processing for package analysis and data insertion has minor and nonexistent overheads, respectively. Finding inconsistencies by running queries from the database has some overhead. This overhead will increase as more tools and data are inserted into the database. This is why it might be a good idea to use Machine Learning for dynamic classification. There is medium overhead when training the data but almost no overhead when performing classification.

### IX. DISCUSSION AND LIMITATIONS

In this section, we discuss the limitations of our methodologies and the challenges we faced in implementing our framework. We also highlight future research directions, motivated by our findings.

**Tool Dependency and Complexity**. Despite the fact that we implemented an extensible framework for reducing inconsistencies for different severity ratings, our approach is still totally dependent on the tools themselves. Consequently, if the tools do not standardize their results or do not maintain quality in their product, the performance of our framework might fall. In addition, not all the tools are easy to set up and use. Even though our framework is extensible for any tool, learning about the tool itself, might not be straightforward.

**Resolving Mismatches Bottom Up**. To prevent resolving the same CVE more than once, we chose to go with the fix that had the most recent 'Modification Time' along with the greater number of matching pairs. Although choosing the most recent data seemed obvious, this led to the exclusion of all other inconsistencies for that CVE from the rest of the dataset. That is, if we solve an inconsistency with the aforementioned approaches once, we do not solve it again, even though it may have existing records for inconsistent assigners, package names, and versions. This may lead to minor errors in vulnerability labeling.

**Manual Label Generation**. Our goal in LUCID is to automate the process as much as possible. Even though we use different Machine Learning Algorithms in our multi-classification problem, the generation of the dataset required manual effort. Also, resolving the inconsistencies took additional manual effort, like filtering, matching, querying, etc.

**Limited Number of Algorithms Used**. Since there hasn't been much work done, for solving inconsistencies in severity ratings for containers, further evaluation with more algorithms is required. We believe the current accuracy of 84% and F1-Score of 83% could be improved.

**Basic Neural Network**. Due to the lack of time, we only used one neural network algorithm in our Dynamic Approach, which was the simplest one. This is probably the reason why it performed badly in one of the classes. More fine-tuning is needed by the neural network so that it can correctly classify the data for all the classes, even the ones with class imbalance.

## X. CONCLUSION AND FUTURE WORK

Different container scanning tool returns different results scanning the same sets of images. This causes confusion among individuals, and companies especially cloud vendors, regarding the security of their container subsystems. Our work explores finding the causes of inconsistencies among different scanning tools and resolving them with the help of LUCID. We focus on finding the sources for the inconsistencies, including inconsistent package names and versions, assigners, and modification times. The category with inconsistent assigners had the highest percentage of inconsistency, which was 95%. We used two approaches, namely the recent and voting approach to reduce the number of inconsistencies and generate the correct label for a vulnerability. In fact, we were able to reduce 70% of the inconsistencies encountered, from an average of 9.98 to 2.89 inconsistencies per image.

In addition, we also implemented dynamic classification of the severities using popular machine learning models. We found that Decision Tree performs best in classification, followed closely by Random Forest, although the latter proved to be more resource-efficient.

Even though there has been a limited amount of research in the area for reducing inconsistencies, to the best of our knowledge, this work is a first of its kind that focuses on inconsistencies among containers, scanning tools and provides remediation. Individuals and cloud providers can use our framework to secure their container systems. They can also include their modified 'Environmental Metric' in LUCID.

Future work includes automated Label Generation with intuitive Self-Learning Models and investigation of fine-tuned machine learning and CNN models to increase the classification rate of the framework. Methods that can prove the credibility of less popular assigners can lead to the inclusion of more data, which would further reduce the inconsistencies. We also plan to make our dataset public for others to use.

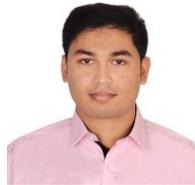

**Md Sadun Haq** received his B.Sc. degree in Computer Science and Engineering from Dhaka University in 2017. He is currently pursuing a Ph.D. degree in Computer Science at The University of Texas at San Antonio. He is currently working as a Graduate Assistant in the Computer Science Department since 2019. His research interests include the security of Container Technology, the Internet of Things, Edge Computing, Privacy Policy, and Anomaly Detection among these technologies.

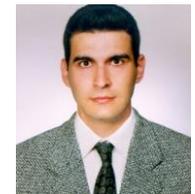

**Ali Saman Tosun** received his BS degree in 1995 and MS degree in 1997 computer engineering from Bilkent University, Ankara, Turkey. He received MS and Ph.D. degrees in computer science and engineering from the Ohio State University, in 1998, and 2003, respectively. He joined Department of Computer Science, the University of Texas at San Antonio in 2003 as an assistant professor and worked there till 2021. Currently, he is Allen C. Meadors Endowed Chair in Computer Science at the University of North Carolina at Pembroke. His research interests include network security, software-defined networks, Internet Of Things, storage systems, and large-scale data management.

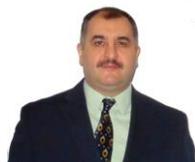

**Turgay Korkmaz** received the B.Sc. degree (Hons.) in computer science and engineering from Hacettepe University, Ankara, Turkey, in 1994, the M.Sc. degree in computer engineering from Bilkent University, Ankara, the M.Sc. degree in computer and information science from Syracuse University, Syracuse, NY, USA, in 1996 and 1997, respectively, and the Ph.D. degree in electrical and computer engineering from the University of Arizona, in December 2001, under the supervision of Dr. M. Krunz. In January 2002, he joined The University of Texas at San Antonio, as an Assistant Professor with the Computer Science Department. He is currently a Full Professor with the Computer Science Department. He is also involved in the area of computer networks, network security, network measurement and modeling, and Internet related technologies. His research interests include quality-of-services (QoS)-based networking issues in both wireline and wireless networks.